\pdfoutput=1

\documentclass[a4paper,referee]{raa}            

\usepackage{graphicx,times}  

\usepackage{natbib}
\usepackage{booktabs}
\usepackage{amssymb,amsmath}
\bibpunct{(}{)}{;}{a}{}{,}
\usepackage{lineno}
\usepackage{CJK}
\usepackage[pagebackref=true,colorlinks,linkcolor=blue,anchorcolor=blue,citecolor=blue]{hyperref}

\begin{document}

  \title{The Mini-SiTian Array: Imaging Processing Pipeline}

   \volnopage{Vol.0 (20xx) No.0, 000--000}    
   \setcounter{page}{1}   

   \author{Kai Xiao 
      \inst{1}
      \and Zhirui Li \footnote[4]{Kai Xiao and Zhirui Li contributed equally to this manuscript.}
      \inst{2,1}
      \and Yang Huang
      \inst{1,2,3,4}
      \and Jie Zheng
      \inst{3}
      \and Haibo Yuan
      \inst{4,5}
           \and Junju Du
      \inst{6}
      \and Linying Mi
      \inst{3, 1}
      \and Hongrui Gu
      \inst{3,1}
      \and Yongkang Sun
      \inst{3,1}
    \and Bowen Zhang
      \inst{3,1}
      \and Shunxuan He
      \inst{3,1}
      \and Henggeng Han
      \inst{3}
      \and Min He
      \inst{3}
    \and Ruifeng Shi
      \inst{1}
      \and Yu Zhang
      \inst{3}
      \and Chuanjie Zheng
      \inst{3,1}
      \and Zexi Niu
      \inst{1}
      \and Guiting Tian
      \inst{1}
            \and Hu Zou
      \inst{3}
            \and Yongna Mao
      \inst{3}
   \and{Hong Wu}
      \inst{3}
   \and{Jifeng Liu}
      \inst{2, 3, 1, 4}
   }

   \institute{School of Astronomy and Space Science, University of Chinese Academy of Sciences, Beijing 100049, China;
        \and
    New Cornerstone Science Laboratory, National Astronomical Observatories, Chinese Academy of Scciences, Beijing, 100012, China; 
       \and
    National Astronomical Observatories, Chinese Academy of Sciences,
             Beijing 100012, China;
       \and
           Institute for Frontiers in Astronomy and Astrophysics, Beijing Normal University, Beijing, 102206, China;
        \and
    School of Physics and Astronomy, Beijing Normal University No.19, Xinjiekouwai St, Haidian District, Beijing, 100875, China;
       \and
    Shandong Key Laboratory of Optical Astronomy and Solar-Terrestrial Environment, School of Space Science and Physics, Institute of Space Sciences, Shandong University,  Weihai, Shandong, 264209, China.
      \\
    Corresponding author: Yang Huang (huangyang@ucas.ac.cn)\\
\vs\no
   {\small accepted 11-Feb-2025}}

\abstract{As a pathfinder of the SiTian project, the Mini-SiTian (MST) array, employed three commercial CMOS cameras, represents a next-generation, cost-effective optical time-domain survey project. This paper focuses primarily on the precise data processing pipeline designed for wide-field, CMOS-based devices, including the removal of instrumental effects, astrometry, photometry, and flux calibration. When applying this pipeline to approximately 3000 observations taken in the Field 02 (f02) region by MST, the results demonstrate a remarkable astrometric precision of approximately 70--80\,mas (about 0.1\,pixel), an impressive calibration accuracy of approximately 1\,mmag in the MST zero points, and a photometric accuracy of about 4\,mmag for bright stars. Our studies demonstrate that MST CMOS can achieve photometric accuracy comparable to that of CCDs, highlighting the feasibility of large-scale CMOS-based optical time-domain surveys and their potential applications for cost optimization in future large-scale time-domain surveys, like the SiTian project.
}

   \authorrunning{Xiao et al.}       
   \titlerunning{The Mini-SiTian Survey: Data Accurate Processing Pipeline}  

   \maketitle

\section{Introduction}
Continuous observations of the sky have provided fascinating insights into the dynamic universe. From Hipparchus's detection of the precession of the equinoxes to Halley's discovery of stellar proper motion \citep{2009AAS...21332001B}, the cataloging of over thousands of candidate exoplanets by Kepler \citep{2010Sci...327..977B} and the Transiting Exoplanet Survey Satellite (TESS; \citealt{2015JATIS...1a4003R}), and a surge of researches on various types of transient sources has been sparked by the Zwicky Transient Facility (ZTF; \citealt{Bellm_2018}), the various fields within astronomy have achieved significant advancements.

Although time-domain surveys such as TESS and ZTF provide valuable monitoring capabilities, continuous observations over a larger region of the sky with higher cadences and greater depth will be necessary to further advance our understanding of the universe.
Furthermore, improving the cost-effectiveness of surveys is essential in the era of large-scale observations, especially for large-scale telescope arrays. Over the past few decades, Charge-Coupled Devices (CCDs) have been widely used in large-scale time-domain surveys.
The most significant differences between CCDs and Metal-Oxide-Semiconductor (CMOS) sensors lie in how they process each pixel. CCD sensors read out the data from all pixels uniformly, while CMOS independently converts photons to electrical signals for each pixel. The manufacturing process for CMOS sensors is similar to standard microelectronic integrated circuit fabrication, which results in lower power consumption, faster speed, and lower cost.
Following these lines, Tsinghua University-Ma Huateng Telescopes for Survey \citep[TMTS;][]{2020PASP..132l5001Z} and the Large Array Survey Telescope \citep[LAST;][]{2023PASP..135f5001O} were conducted.
Their data processing resembled differential photometry, achieving an accuracy of several milli-magnitudes; however, the generalized absolute photometric calibration is often of greater importance and receives more attention in CMOS-based surveys.

As a next-generation time-domain survey, the SiTian project, equipped with CMOS detectors, will cover a large sky area ($\ge$~10,000 square degrees) across $gri$ three bands, with a 30-minute cadence, reaching a limiting magnitude of 21 \citep{2021AnABC..93..628L}. To achieve the scientific goals of the SiTian project, the high-precision processing of photometric data from CMOS detectors is crucial. On the one hand, CMOS offers a low-power, cost-effective alternative and has nearly matched the performance of CCDs in electronic instability, as described in \cite{2023PASP..135e5001A} and Zhang et al. (submitted to this volume). On the other hand, for high-precision photometry, CMOS still faces certain limitations, such as the roll shutter’s unsuitability for ultra-fast exposures and the ``pepper \& salt'' (PS) effect \citep{2023PASP..135e5001A}. The PS effect, a unique and challenging issue, is specific to CMOS detectors and arises due to their manufacturing process. Once the advantages of CMOS in time-domain surveys are validated, it is poised to become a powerful tool, significantly reducing project costs and advancing modern astronomy.

As a pathfinder project, the Mini-SiTian (MST) at the Xinglong observation station houses three 30\,cm telescopes equipped with a commercial-grade CMOS detector with $9\rm k$ $\times$ $6\rm k$ pixels, covering a 3\,deg$^2$ field of view and yielding a 0.86 pixel$^{-1}$ pixel scale. The MST aims to search for transit signals of exoplanets, optical counterparts of gravitational waves, supernovae, fast transients, etc. Detailed science goals are outlined in H. Han (to be submitted). Since 2022, the MST has conducted multiple repeated observations of several specific regions using $G$/$g$, $R$/$r$ and $i$-filters installed on the MST2, MST3 and MST1 telescopes, respectively.

This paper aims to describe the data acquisition and storage processes of MST, with a particular focus on the development of a high-precision photometric pipeline that achieves a photometric accuracy of less than 1\%. This paper is organized as follows: In Section\,\ref{sec1}, we provide a brief overview of MST's hardware system and software control, as well as the construction of the data processing pipeline presented in Section\,\ref{sec2}. Section\,\ref{sec3} analyzes and discusses the processing results of sample images taken during the pilot phase. Finally, a summary is given in Section\,\ref{secsum}.

\section{Hardware System and Software Control: A Brief Overview} \label{sec1}
\subsection{Enclosure and Telescope}
The MST array, located at the Xinglong Observatory, consists of three telescopes (MST1, MST2 and MST3) with identical refractive optical configurations (He et al. submitted to this volume). Each telescope features a 300\,mm primary mirror, operates at a focal ratio of $f/3$, and is equipped with a camera positioned at the Cassegrain focus. The mean and median values of seeing at Xinglong observatory are around $1.9^{\prime\prime}$ and $1.7^{\prime\prime}$ \citep{2015PASP..127.1292Z}, respectively. For more information about the site and telescope system, please refer to He et al. (submitted to this volume).

\subsection{Filters}
The three telescopes (MST2, MST3 and MST1) are equipped with SDSS-like $gri$-filters, respectively, enabling the simultaneous capture of three-band photometric data for a given sky region. The central wavelengths of the $gri$-filters are approximately 4800, 6200 and 7700\,\AA, respectively, with corresponding full width at half maximum (FWHM) being approximately 1200, 1500 and 1300\,\AA. 

In the early stages of observation, since the $gr$-filters had not yet been prepared, MST1 and MST2 were temporarily fitted with $G$- and $R$-filters for tests. The central wavelengths of the $GR$-filters are approximately 5200 and 6500\,\AA, respectively, with corresponding FWHM being approximately 800\,\AA~each. For example, the entire Field 01 and part of Field 02 sky regions were imaged using $G$- and $R$- filters. The transmission curves for $gGrRi$-filters are shown in Fig.\,1 of Xiao et al. (to be submitted) and also He et al. (submitted to this volume).

\subsection{Detector}
Each MST telescope is equipped with the ZWO ASI6200MM Pro CMOS detector, which features a commercial-grade Sony IMX411 CMOS sensor chip that inherently supports native 16-bit depth raw image output. This detector has a resolution of $9576\times 6388$ pixels, with a pixel size of $3.76$ $\mu$m corresponding to $0.862^{\prime\prime}$. The quantum efficiency of the detector can be found in Fig.\,1 of Xiao et al. (to be submitted). The typical read noise of the detector is about 1.028 $e^-$ at a gain of 0.25 $e^-\,{\rm ADU^{-1}}$, and the dark current is about 0.002 $e^-\cdot {\rm pixel}^{-1}\cdot {\rm s^{-1}}$ at $0\,^{\circ}\text{C}$ (Zhang et al. submitted to this volume).

When operating, the CMOS sensor is cooled to 30 degrees Celsius below the ambient temperature. Using the ASCOM protocol, the cameras facilitate the transfer of raw image data in the standard FITS format. After capturing the images, the MST camera transmits the 16-bit depth raw image data to a nearby micro-host via a USB cable. Subsequently, this micro-host transfers the raw data to the MST server through a local wireless network. It is important to note that the raw data remains uncompressed throughout the transfer process, ensuring data integrity. 

\begin{figure}
    \centering
    \includegraphics[width=14cm]{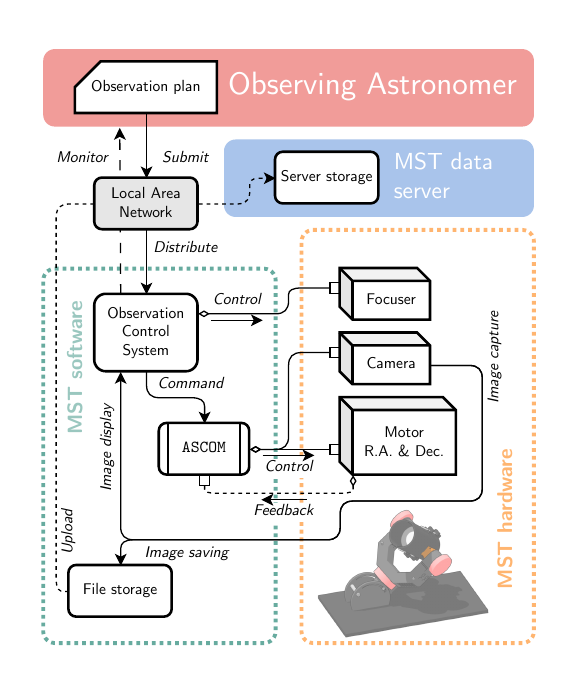}
    \caption{The schematic diagram of the MST control system, data acquisition, and storage. The MST observing facility comprises hardware and software components interconnected via a USB cable. The software, integrated into a PC linked to the telescope, controls the hardware's electronic focusing system via the Observation Control System (OCS). It also leverages the {\tt ASCOM} software package to manage the camera (including cooling and exposure) and telescope motors, fulfilling observational requirements. Real-time feedback to the OCS allows monitoring of hardware status. Prior to observations, astronomers remotely submit plans to the OCS through Local Area Network (LAN). Subsequently, the OCS executes these plans, relaying hardware status and acquired images back to the astronomer over the same network. These images are locally stored and forwarded to the MST server via the LAN for further processing.}
    \label{fig:MST_control}
\end{figure}

\subsection{Software Control}
The schematic diagram of the MST software control, data acquisition, and storage is shown in Fig.\,\ref{fig:MST_control}. Each telescope in the MST array is connected to a standalone control computer (MST mini host), allowing the controller to connect to the host via a wireless router. Once connected, the controller can generate ASCOM commands using the Observation Control Software (OCS) on the MST host to operate the telescope for observations. Status information from the telescope (including pointing, focus position, CMOS image, etc.) is also fed to the OCS in real-time to help the controller evaluate the telescope's operational status. 

Exposure images are stored on the MST host and uploaded to the MST server through the wireless router for subsequent data processing once the observation is completed. In addition to the components controlled by the software, observations also require calibration, which includes acquiring FLAT, DARK, and BIAS frames. These calibrations necessitate manual intervention by the controller, such as installing flat field plates and dust caps. The calibration frames are also uploaded to the MST server at the beginning and end of the observation. Detailed information can be found in He et al. (submitted to this volume).

\begin{figure}
    \centering
    \resizebox{\hsize}{!}{\includegraphics{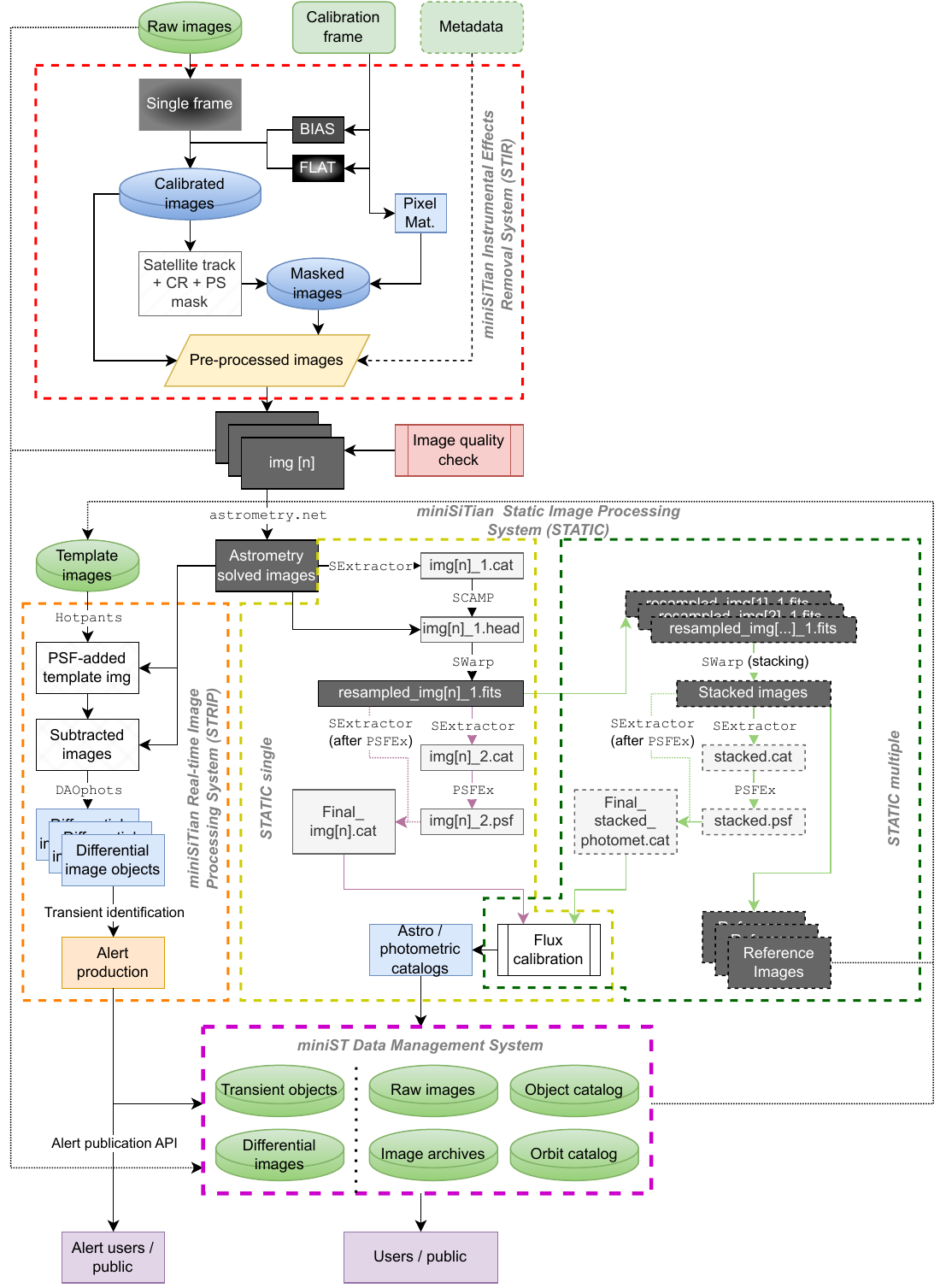}}
    \caption{Flowchart of the MST data processing pipeline.}
    \label{fig:pipeline}
\end{figure}

\begin{figure}
    \centering
    \resizebox{\hsize}{!}{\includegraphics{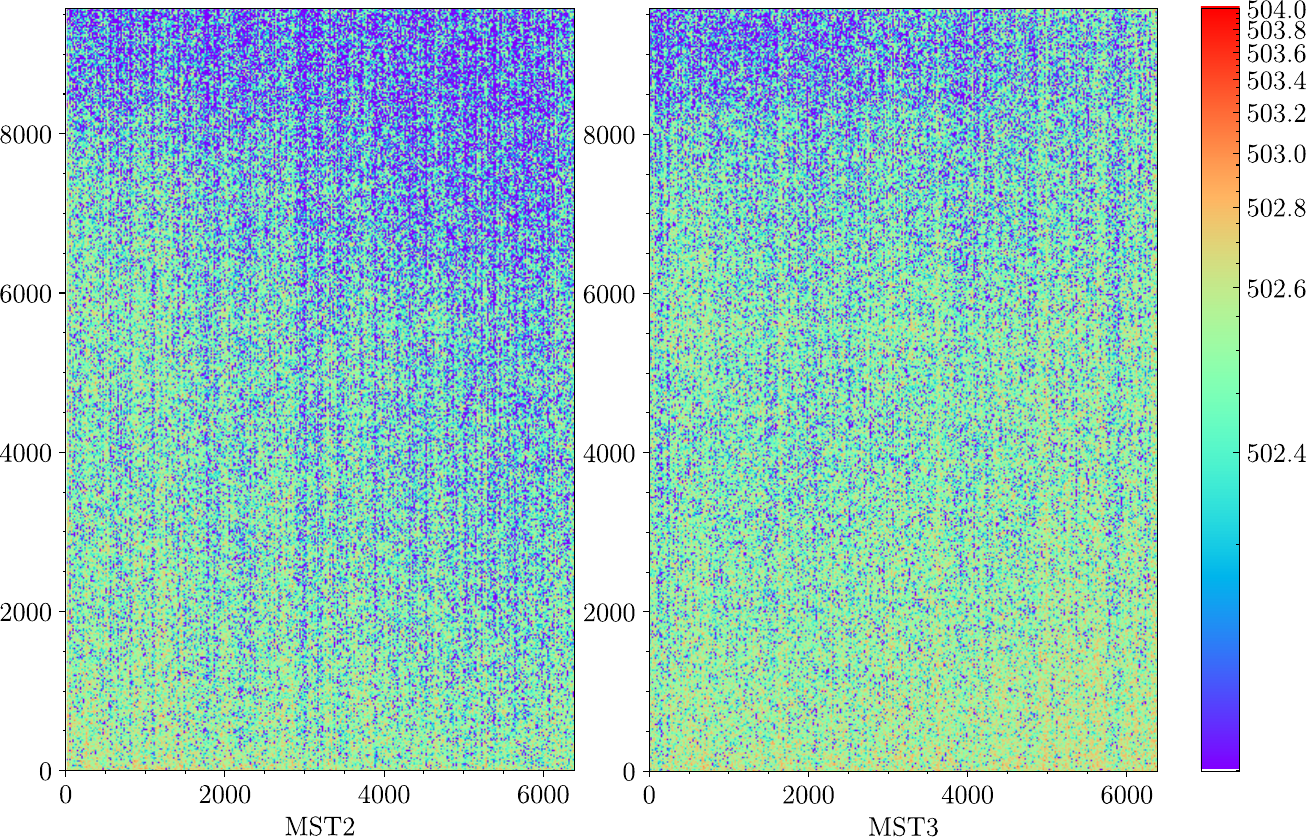}}
    \caption{An example to illustrate a MST2 (the left panel) and MST3 (the right panel) combined master bias frames collected on January 18, 2023. Due to the limited number of common observations between MST1 and the other two telescopes, we do not present MST1's results here.}
    \label{fig:bais}
\end{figure}

\begin{figure}
    \centering
    \resizebox{\hsize}{!}{\includegraphics{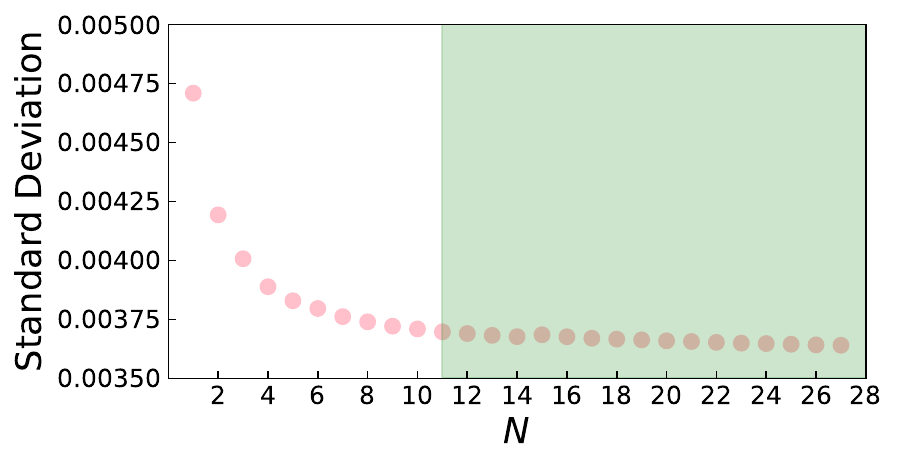}}
    \caption{Taking the observation by MST2 in January 2023 as an example to show the variation of the standard deviation of small-scale flat with the number of nights ($N$). When $N$ exceeds 11 (green area), the standard deviation of the small-scale flats stabilizes, primarily contributed by pixel-to-pixel variations.}
    \label{fig:sig_n}
\end{figure}

\begin{figure}
    \centering
    \resizebox{\hsize}{!}{\includegraphics{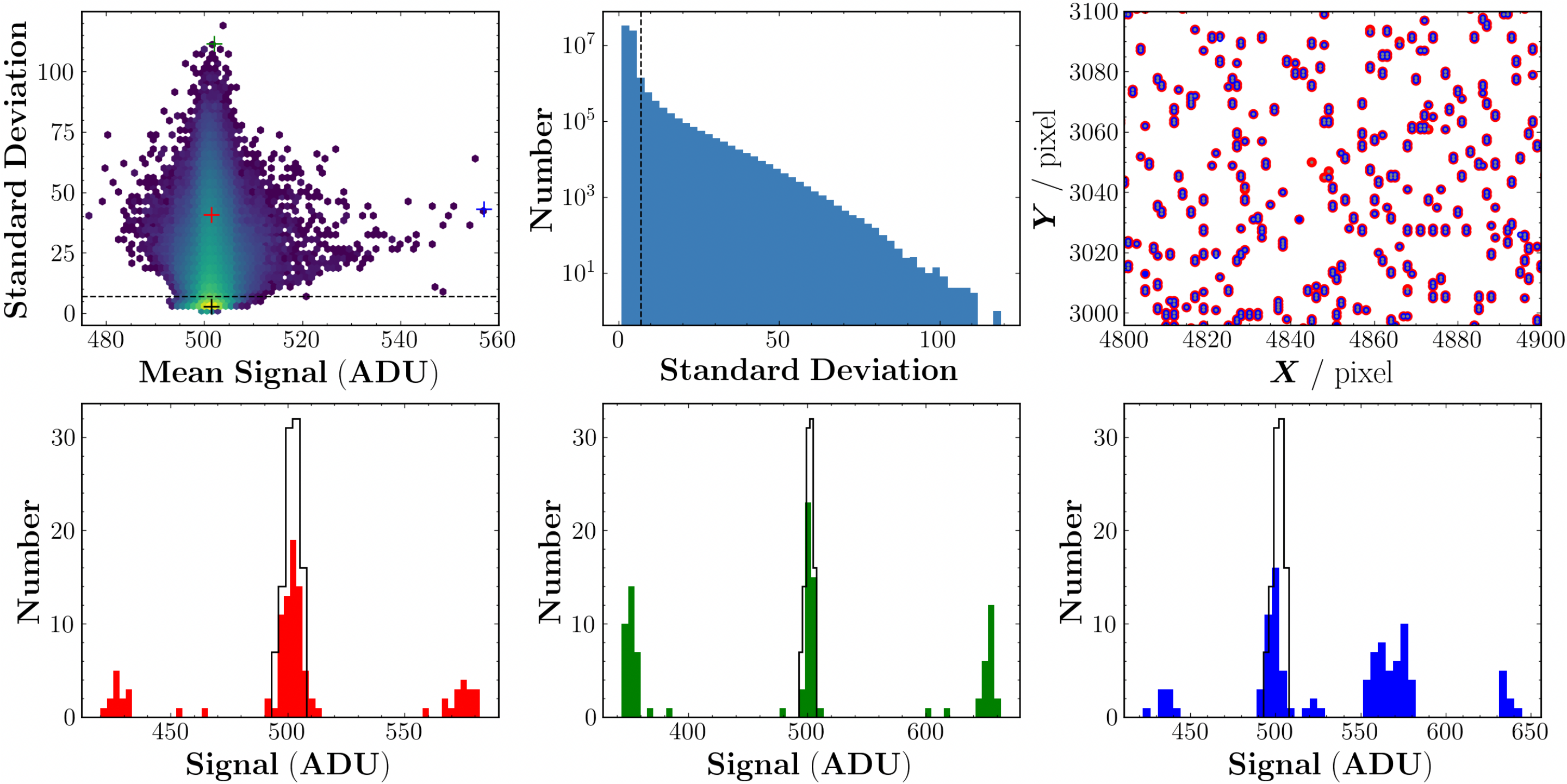}}
    \caption{The characteristics of ``pepper \& salt'' (PS) noise using MST2 observations as an example. The top left panel: The consistency of ADU values at corresponding pixels over 100 bias frames as a function of ADU. The colored plus signs represent randomly selected pixels at various locations, and the black represent a normal pixel. The top middle panel: The histogram distribution of the consistency of ADU values at corresponding pixels over 100 bias frames. The black dashed line roughly indicates the boundary between normal pixels and PS pixels, which is also shown in the top left panel in spatial space. The top right panel: The distribution of PS pixels in the sub-detector region. The bottom panels: The histogram distributions of ADU values across 100 bias for the three pixels marked with colored plus signs in the top left panel. For each bottom panel, the black histogram represent the distributions of ADU values of the normal pixel.}
    \label{fig:ps}
\end{figure}

\section{Imaging processing pipeline} \label{sec2}
The MST imaging processing system is divided into three main components: the data storage/management system, the data processing system, and the data products. For a description of the data storage/management system, please refer to He et al. (submitted to this volume). The data processing system comprises real-time image subtraction (Gu et al., submitted to this volume), machine learning-based classification of variable sources (Shi et al., to be submitted), and a high-precision data processing pipeline (this work).
The real-time processing results are made available as real-time data products, with detailed information on the transient source provided by Gu et al. (submitted to this volume) and Shi et al. (to be submitted), and the planet catalog managed by Liu et al. (submitted to this volume).
The high-precision data processing results are offered as survey data products, including images and catalogs, which will be released soon as MST DR1 (Mi et al., in prep.) and can be accessed on the website (\href{https://nadc.china-vo.org/sitian/}{https://nadc.china-vo.org/sitian/}).

This section introduces the high-precision data processing pipeline for MST observations, the focus of this paper, which includes the MST Instrumental Effects Removal System (STIR), the MST Static Image Processing System (STATIC), and the MST Real-Time Image Processing System (STRIP; Gu et al. submitted to this volume), as shown in Fig.\,\ref{fig:pipeline}. 

\subsection{STIR}
As shown in Fig.\,\ref{fig:pipeline}, STIR takes a single frame of raw images, along with calibration frames including bias and flat frames, as input, and outputs images corrected for bias and flat, identified and marked with instrumental artifacts such as bad pixels, saturated pixels, PS pixels, cosmic rays, and satellite trajectories. Given that the CMOS camera employed by the MST exhibits an exceptionally low dark current, the effect of dark has not been considered within the STIR process. We only selected the pixels that fall outside $3\sigma$ in the dark frames as bad pixels.

\subsubsection{Bias Subtraction}
At the beginning of each night's MST observations, 10 bias frames with zero exposure were collected. First, we combine all ten bias frames to generate a master bias for that night. An example of the combined bias frame from MST2 and MST3 on January 18, 2023, is shown in Fig.\,\ref{fig:bais}. The combined frames reveal several notable patterns, which have been carefully analyzed in Zhang et al. (submitted to this volume). Then, for each night, we subtract the master bias from all the observed images and flat frames to obtain the bias-subtracted frames.

\subsubsection{Flat Correction}
At the beginning of each night's MST observations, approximately 20 twilight flat-fielding frames with an average value of 20,000 ADU were collected. To generate the master flat, we apply median smoothing to each bias-subtracted flat using a 51-pixel window, resulting in a large-scale flat field. By dividing the original bias-subtracted flat by the large-scale flat, we obtain a small-scale flat field. We then combine all 20 small-scale flats for each night to generate a master flat for that night. The signal-to-noise ratio (SNR) of the master flat is approximately $\sqrt{20000\times 0.25\times 20}\simeq 313$, with a typical pixel ADU value of 20,000 and a gain of 0.25 $e^{-}\,{\rm ADU^{-1}}$. To ensure flat-field accuracy better than one thousandth, we combine the flat-field frames from the five consecutive days before and after, as well as the flat-field frame from the current day, to create the master flat frame for the current day. If the flat-field frames used span over two months, any flat-field frames beyond this two-month span will not be utilized.

The operable assumption mentioned above is that the small-scale flat fields remain stable over the month, which has been confirmed in the work of Hu et al. (in prep.). Hu et al. (in prep.) first calculated the standard deviation of the stacked flat fields at different days within a month. They found that as the number of adjacent days increases, the standard deviation of the small-scale flat field after stacking decreases rapidly and tends to stabilize when the number of days exceeds 11, as shown in Fig.\,\ref{fig:sig_n}. When $N$ is greater than 11, the random errors in small-scale flats can be negligible, and at this point, the scatter of small-scale flats is almost dominated by pixel-to-pixel non-uniformity. Using a master flat field obtained by stacking small-scale flat fields over 11 days can effectively improve the flat-field correction precision to less than one-thousandth. For a more detailed description, please refer to Hu et al. (in preparation).

\subsubsection{Definition of PS Pixels}
PS effect refers to a type of noise arising from the unique manufacturing processes of CMOS sensors, characterized by a pattern that resembles ``salt'' and ``pepper'' in the image. To identify pixels affected by PS noise, we collected 100 bias frames using MST2 and MST3 on a single night, respectively. We then plotted the distribution of ADU values for each pixel across these 100 observations, which exhibited a badminton-like distribution, as shown in the top-left panel of Fig.\,\ref{fig:ps}. Most pixels had a standard deviation of ADU values below 8 (a typical value), while only a small fraction (less than 8\%) displayed higher standard deviations, with some exceeding 100 ADU, as clearly illustrated in the top-middle panel of Fig.\,\ref{fig:ps}. In the current pipeline, pixels with an ADU standard deviation greater than 8 were classified as PS pixels. Here, we make the simplified assumption that the salt-and-pepper pixels remain stable over time, but this will require further checks and tests in future work.

To minimize misclassification, we randomly divided the 100 bias frames into three non-overlapping groups in time. We then defined the intersection of the PS pixels identified in these three groups as the final set of PS pixels. The distribution of PS noise in the detector sub-spaces for the three groups is presented in the top right panel of Fig.\,\ref{fig:ps}. Results from the first group are represented by larger red points, those from the second group by smaller blue points, and those from the third group by even smaller gray points. It can be observed that the measurements of most PS pixels are consistent across the three observations, with this consistency quantified to exceed 90\%. We can also observe that the ADU variation of the PS pixels can be as high as over 150 ADU (see the bottom panels). 

For faint stars with an ADU of approximately 5000 that occupy about 10 pixels on the MST detector, the presence of PS noise—constituting about 8\% of the pixels and exhibiting an ADU variation of typically 150—has an influence of roughly 1.5\% on the ADU. This leads to a magnitude error exceeding 0.03\,mag. In contrast, for bright sources, assuming the star's ADU is 50,000 and it covers 10 pixels, this error decreases to less than 0.003\,mag. Hence, PS noise has a greater influence on faint sources than on bright ones. The above is merely a rough order of magnitude estimation, and we intend to further explore the properties and correction methods of PS noise in the future.

\subsubsection{Identifying Satellite and Aircraft Tracks}
\begin{figure}
    \centering
    \resizebox{\hsize}{!}{\includegraphics{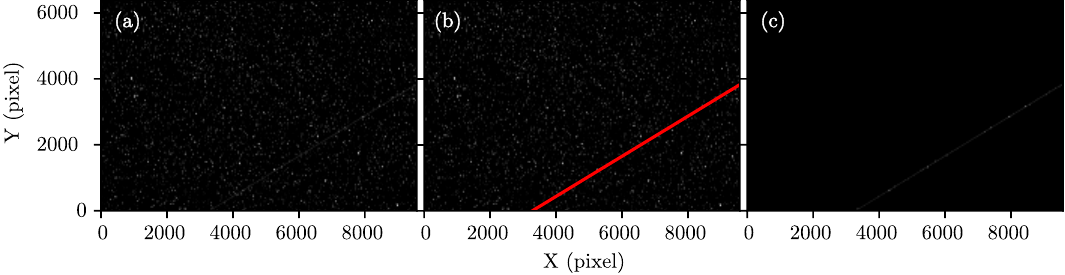}}
    \caption{Taking the observation by MST2 in 18 January 2023 as an example showing the identification of the satellite track. Panel (a): The raw image. Panel (b): Satellite track labeled (red line). Panel (c): Pixels labeled as satellite track in the MASK image.}
    \label{fig:mask}
\end{figure}

This section outlines the identification and labeling of satellite and aircraft tracks.
With the rapid development of commercial aerospace, ground-based optical astronomical observations are increasingly affected by interference from satellite and aircraft trajectories. To address this issue, we designed a specific function in the pipeline to detect satellite trajectories in images. Although MST has a relatively large field of view (FoV), we have found that satellite trajectories are typically represented as clear straight lines in the images collected by MST. Therefore, we employ the Hough transform (\citealt{osti_4746348}) to detect these straight lines. First, we bin the image by a factor of 2 to expedite processing speed, and then we perform the Hough transform using the \texttt{hough\_line} function from the Python package \texttt{skimage.transform} to identify the centerline of the satellite trajectory. Finally, we marked pixels within a 10-pixel width around the centerline that had values exceeding 3 standard deviations above the background median as being affected by the satellite trajectory. 
An example of the satellite track mark process is shown in Fig.\,\ref{fig:mask}.

\subsubsection{Masks}
In this section, we introduce the masks for various types of problematic pixels, including bad pixels, PS pixels, pixels affected by satellite and aircraft trails, and saturated pixels (exceeding 50,000 ADUs). The identification methods for the remaining effects have been described previously. The mask value of each pixel is influenced by various marked pixels, each associated with an integer value of $2^n$, as listed in Table\,\ref{tab:1}. For example, if a pixel in a flag image has a value of 25, it can be expressed as follows: $n=25=1\times2^4+1\times2^3+0\times2^2+0\times2^1+1\times2^0$. This indicates that the pixel exhibits the following defects: cosmic rays, satellite track interference, and bad pixels. 

\begin{table}[h!]
    \centering
    \begin{tabular}{c|c|c|c|c|c} \hline
        Flags & Cosmic Ray & Satellite Track & Pepper \& Salt & Saturated Pixels & Bad Pixels \\
              & ($2^4$) & ($2^3$) & ($2^2$) & ($2^1$) & ($2^0$) \\ \hline
        1 & 0 & 0 & 0 & 0 & 1 \\ 
        2 & 0 & 0 & 0 & 1 & 0 \\ 
        4 & 0 & 0 & 1 & 0 & 0 \\ 
        8 & 0 & 1 & 0 & 0 & 0 \\ 
        16 & 1 & 0 & 0 & 0 & 0 \\ \hline
    \end{tabular}
    \caption{Flags in the MST's Mask and their meanings. From left to right, the binary bit positions start from the second column.}
    \label{tab:1}
\end{table}

\subsection{STATIC}
The STATIC program is designed to perform astrometry, photometry, and flux calibration on images that have been pre-processed in the STIP as seen in Fig.\,\ref{fig:pipeline}. From another perspective, STATIC can be applied to both single images and stacked images.

\subsubsection{Astrometry}
To obtain the celestial coordinates of objects, astrometry is carried out in two steps: obtaining a blind solution and obtaining a precise solution. First, we employed \texttt{astrometry.net} \citep{2010AJ....139.1782L}, a mature software that offers satisfactory solving speed, to obtain the blind astrometric solution. Once the field of view size is specified, \texttt{astrometry.net} delivers satisfactory solution speeds. We use the offline version of \texttt{astrometry.net} on the MST server and select the USNO V4.0 catalog as the reference catalog. Although \texttt{astrometry.net} provides basic field distortion correction, it does not meet our requirements for distorted images. Therefore, we will address the field distortion correction in the next step.

As shown in Fig.\,\ref{fig:pipeline}, the image field distortion involves three steps. In the initial step, we employ $\texttt{SExtractor}$ \citep{1996A&AS..117..393B} to process the image and generate a catalog containing all selected sources ($\texttt{img[n]}$\_$\texttt{1.cat}$). Notably, the celestial coordinates in this catalog retain the distortion information.
In the second step, we align the $X$ and $Y$ coordinates with the initial WCS solution using \texttt{SCAMP} software \citep{2006ASPC..351..112B} to calculate the distortion parameters of the image field, using Gaia DR3 as the reference catalog and a two-dimensional fourth-order polynomial. \texttt{SCAMP} generates a \texttt{.head} file for distortion correction.
In the third step, we merge the astrometrically solved image, the flag image, and the \texttt{.head} file. Using \texttt{SWarp} software \citep{2002ASPC..281..228B}, we resample these components to produce the final output: a single corrected image ($\texttt{resampled}$\_$\texttt{img[n]}$\_$\texttt{1.fits}$).

\begin{figure}
    \centering
    \resizebox{\hsize}{!}{\includegraphics{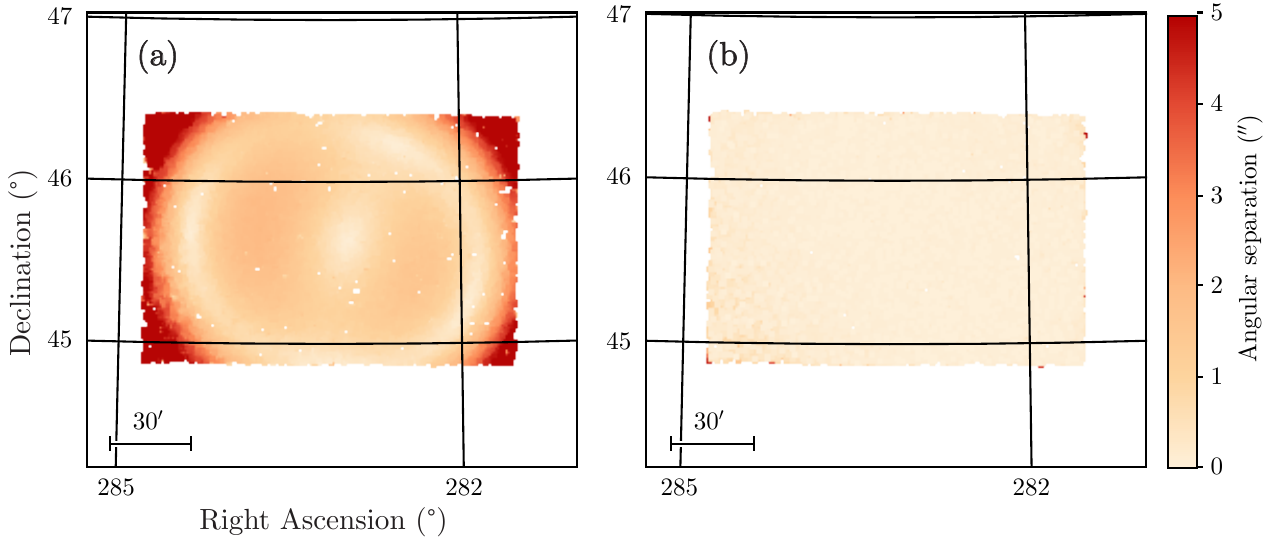}}
    \caption{Taking the observation by MST2 in 18 January 2023 as an example showing the spatial distribution of the cross-matching radius of the cross-matched stars with Gaia DR3 before (left panel) and after (right) astrometry correction in one observation. The color bar, indicated the cross-matching radius with unit of arcseconds, is plotted in the right.}
    \label{fig:ast_corr1}
\end{figure}

\begin{figure}
    \centering
    \resizebox{\hsize}{!}{\includegraphics{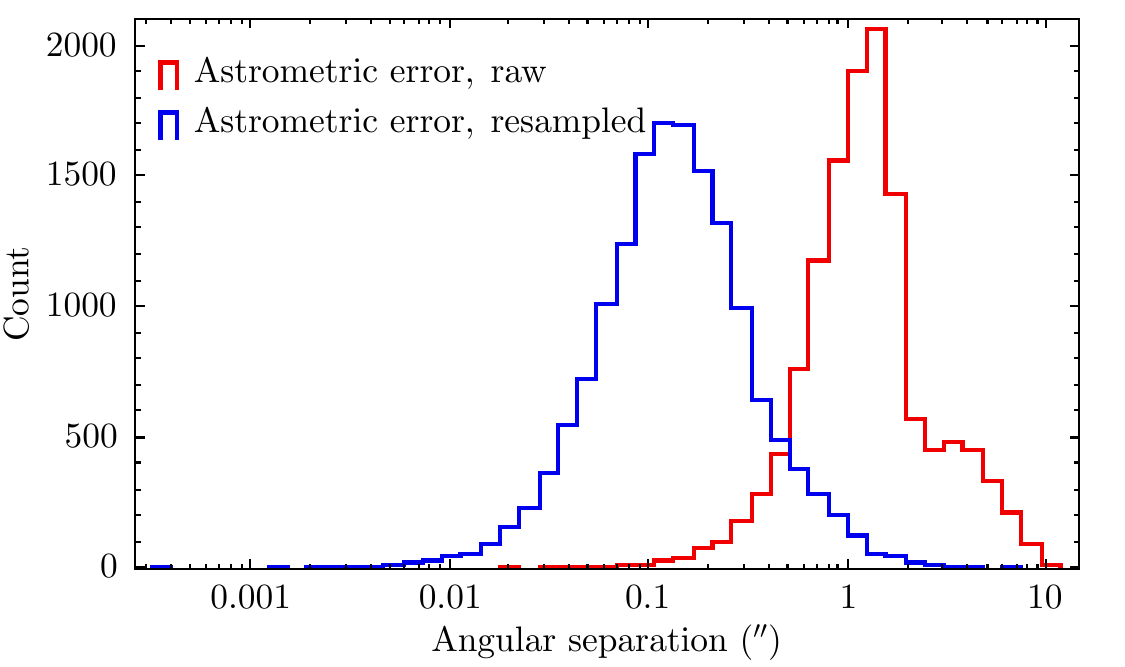}}
    \caption{Taking the observation by MST2 in 18 January 2023 as an example showing the histogram distribution of the cross-matching radius with Gaia data before (red histogram) and after (blue) correction in one observation.}
    \label{fig:ast_corr2}
\end{figure}

The spatial distribution of cross-matched stars with Gaia DR3 before and after distortion correction is shown in Fig.\,\ref{fig:ast_corr1}. The spatial uniformity is significantly improved after the distortion correction. To quantify this improvement, the distribution of astrometric residuals before and after using Gaia data is shown in Fig.\,\ref{fig:ast_corr2}. Essentially, the astrometric residuals with Gaia data decrease from $1 ^{\prime \prime}$ to $0.1 ^{\prime \prime}$, marking a substantial improvement.

\subsubsection{Photometry}
Within the STATIC single pipeline, the distortion-corrected image undergoes two \texttt{SExtractor} processes. In the initial step, the Point Spread Function (PSF) for each star is computed. The resulting catalog, which contains PSF information, is then fed into the PSFEx software \citep{2011ASPC..442..435B}, generating a \texttt{.psf} file. Then, the distortion-corrected image is processed with PSFEx, followed by a third \texttt{SExtractor} operation. This third operation, in conjunction with the \texttt{.psf} file, produces a catalog that includes both aperture and PSF photometry with \texttt{PATTERN\_TYPE=RINGS-HARMONIC}. Additionally, the \texttt{SExtractor} provides results for the BEST aperture magnitude and the Kron magnitude using its default settings. In the MST pipeline, we focus solely on the aperture and the PSF photometry.

\begin{figure*}[ht!] \centering
\resizebox{\hsize}{!}{\includegraphics{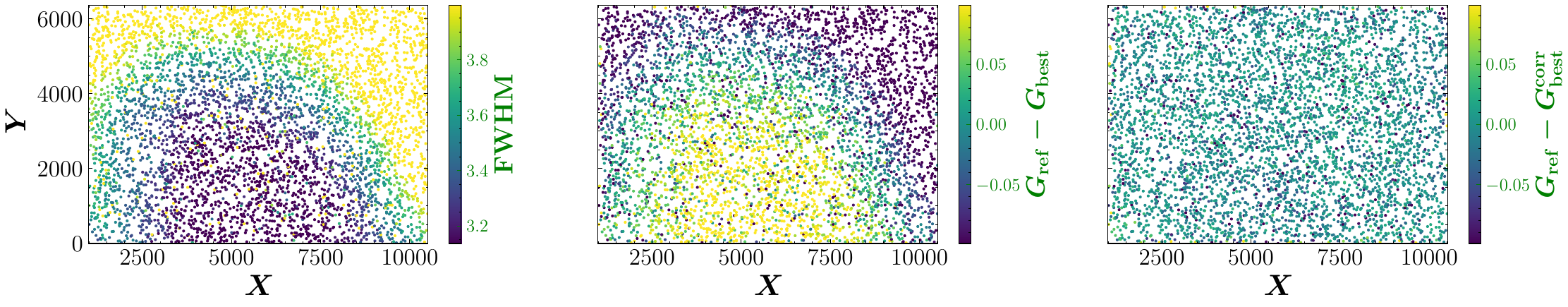}}
\caption{{Taking the observation by MST2 in 18 January 2023 as an example showing aperture correction. Left panel: The distribution of FWHM. Middle panel: The distribution of the difference between the reference aperture magnitude and the best aperture magnitude. Right panel: The residual of the aperture correction.}
}
\label{Fig:apcor}
\end{figure*}

During the aperture photometry process, magnitudes are measured using 25 different apertures that are evenly spaced from 2 pixels to 50 pixels. An excessively large aperture may capture excessive background noise or additional sources, while a too-small aperture might miss necessary source pixels, leading to a reduced signal-to-noise ratio (SNR). Therefore, performing aperture corrections is indispensable to ensure the reliability of the results. Empirically, the best aperture is generally considered to be 1.5--2 times the FWHM \citep{1990PASP..102..932S}. However, for wide-field survey projects, the FWHM varies with the source’s position on the detector (see the left panel of Fig.\,\ref{Fig:apcor}), making it crucial to account for the position-dependent aperture correction. Here, our aperture correction is divided into two steps: first, determining the statistically optimal best aperture, and then considering spatially dependent aperture corrections. 

\begin{figure*}[ht!] \centering
\resizebox{\hsize}{!}{\includegraphics{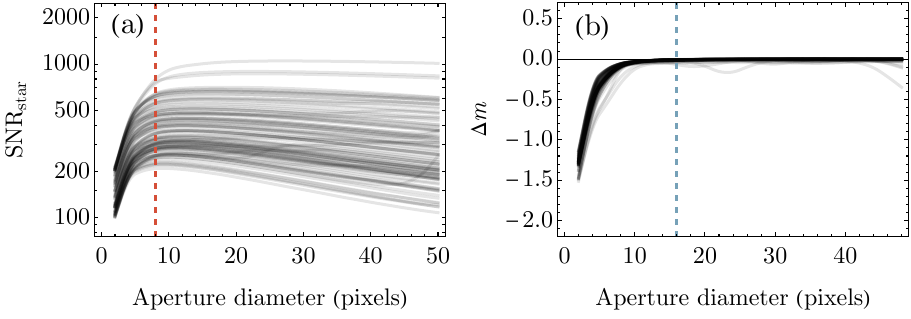}} 
\caption{{Taking the observation by MST2 in 18 January 2023 as an example showing the best and reference aperture size selection. Panel (a): The SNR of 200 randomly selected bright and unsaturated sources at different locations (the four corners and the central region) in the field of view as a function of aperture radius. Panel (b): Growth curves based on the same 200 sources.}
}
\label{Fig:snrrad}
\end{figure*}

To determine the best aperture for each image, we select bright stars with photometric errors ranging from 0.002 to 0.008\,mag as the reference sample. We then plot the dependence of a star's SNR on the aperture for each star. An example of this relationship is shown in panel (a) of Fig.\,\ref{Fig:snrrad}. As observed, the star's SNR quickly rises to its peak value and then slightly decreases as the aperture diameter increases. Finally, the aperture radius that yields the highest SNR for each star is identified, and the median value of these aperture radii for the bright stars in the image is computed as the best aperture value for the image. The best aperture size is approximately 6$^{\prime\prime}$, which is roughly 2 times the FWHM, consistent with empirical expectations. We statistically analyzed all the best aperture sizes from more than 3000 observations and found that 2 times the FWHM is a typical value. Based on this sample, a comparatively larger aperture radius is selected as the reference aperture for computing corrections to the best-aperture magnitudes. To do this, we plot the dependence of the star's magnitude offset, defined as the difference between the magnitude in aperture $n$ and the magnitude in aperture $n-2$, against the aperture diameter $n$. As observed, the star's magnitude offset quickly grows to become stable with increasing aperture diameter, provided the diameter is not excessively large. The aperture radius is chosen when the magnitude offset between aperture $n$ and $n-2$ is less than $1\times 10^{-4}$\,mag for each star. To obtain aperture corrections for all sources, we fit a second-order two-dimensional polynomial to the magnitude differences of the reference sample based on their $X$ and $Y$ coordinates for each image (see the middle panel of Fig.\,\ref{Fig:apcor}). This resulting polynomial is then applied to all sources in the image, and the residual of polynomial fitting is shown in the right panel of Fig.\,\ref{Fig:apcor}.

The STATIC multiple module, on the other hand, utilizes \texttt{SWarp} for the stacking function of multiple images. Given that the MST program involves multiple exposures of the same region of the sky, STATIC single produces a large number of corrected images of the same region, which are stacked by \texttt{SWarp} to produce images with better SNR than a single image. The astrometry, photometry, and flux calibration of the stacked images is similar to that of STATIC single, but with significantly better photometric accuracy, as described in a subsequent section. Additionally, image stacking can deepen the limiting magnitude. The implementation of this process can be divided into two parts: using \texttt{SWarp} for image stacking, followed by processing each individual stacked image through a dedicated pipeline. Further details on this distribution can be available in Xiao et al. (in prep.), and a discussion can be found in Section\,\ref{merging}. 

\subsubsection{Flux Calibration} \label{sec:cal}
A uniform and accurate photometric calibration plays a key role in wide-field surveys. For the photometric calibration of MST, we first derived the transmission curve of the MST system by considering the significant impacts of the Earth's atmosphere, filters, and the CMOS detector. We then convolved the ``corrected'' Gaia BP/RP (XP) spectra \citep[][]{2024ApJS..271...13H}, with the MST's transmission functions, resulting in the apparent magnitudes of approximately 200 million sources in AB system in the MST $gGrRi$-bands. This yielded a standard star library comprising 200 million sources with $gGrRi$ magnitudes, serving as the reference star catalog for MST data photometric calibration.

In the photometric calibration process, both aperture magnitudes after aperture correction and PSF magnitudes for all stars with photometric error less than 0.02\,mag in each image were cross-matched with the standard star library with a cross-radius of 1$^{\prime\prime}$. The cross-matched samples are used as calibration samples to derive the calibration zero point for each image, where the calibration zero point is a function of the source position on the detector. For more comprehensive details and in-depth insights, please refer to the forthcoming work by Xiao et al. (to be submitted).

The MST imaging processing pipeline is currently deployed on a workstation equipped with two Intel 6448 processors, 503 GB of memory, and a 64 TB mechanical hard drive. In addition to the performance of the workstation, the runtime is also directly proportional to the source density within the field of view. Taking f02 as an example, which contains more than 56,000 sources, we provide typical computation times. The pipeline processing time breakdown is as follows: the STIR process takes about 310 seconds for all the observational data from each day, astrometry and photometry take approximately 124 seconds per exposure, and aperture magnitude correction and photometric calibration take around 10 seconds per exposure. For example, processing 79 exposures from January 18, 2023, takes a total of about 10,894 seconds (approximately 3 hours).

\section{Performance and Results}\label{sec3}
Over the past two years, MST has conducted multiple continuous observations of various sky regions, including the f02 region, using MST2 and MST3. In this section, we present the performance of the data processing pipeline using nearly 40 days of continuous observation data from the f02 region acquired with MST2. The demonstration will include astrometric precision, flux calibration precision, photometric accuracy, and the presentation of results for variable sources and supernovae, as well as a simple test of image stacking and catalog merging.

\begin{figure}
    \centering
     \includegraphics[width=13cm]{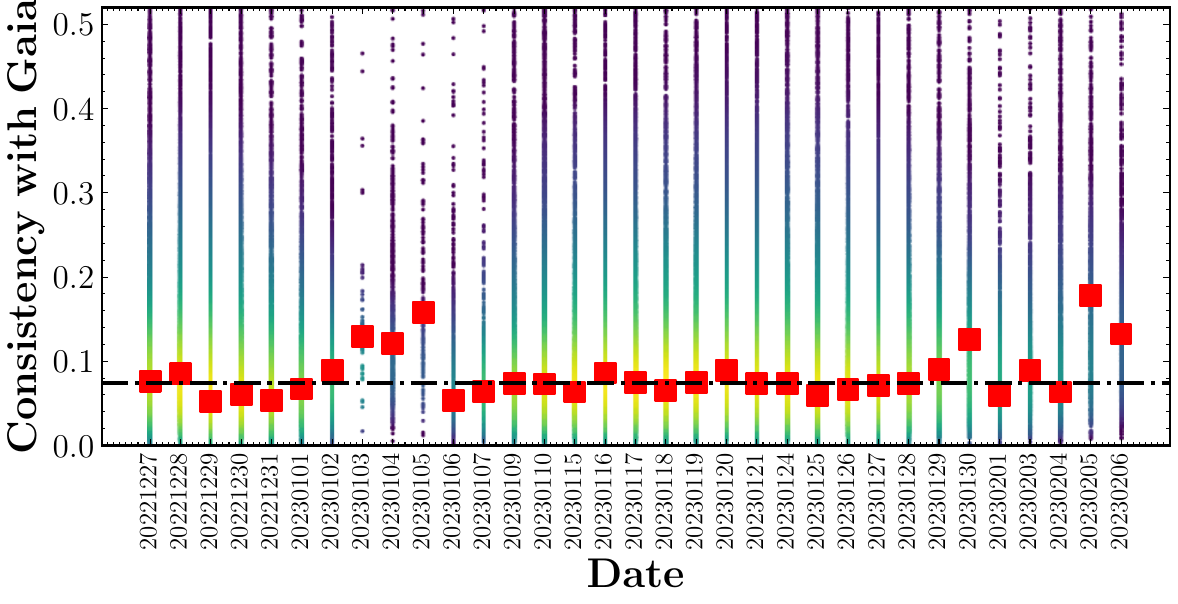}
    \caption{Taking the observation by MST2 in January 2023 as an example showing the astrometric residuals along two axes. The red points represent the median residuals for each day, and the black line represents the overall median value of these red points. Due to unfavorable weather conditions, the number of detected sources on \texttt{20230103}, \texttt{20230104}, \texttt{20230105}, and \texttt{20230205} was relatively low, leading to greater uncertainty in the astrometric precision.}
    \label{fig:astro3}
\end{figure}
\begin{figure}
    \centering
    \includegraphics[width=13cm]{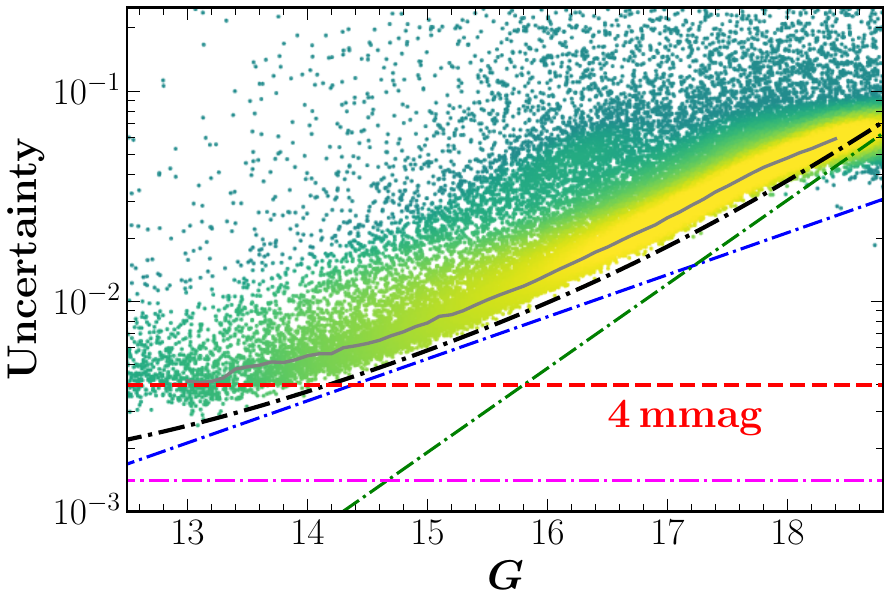}
    \caption{The standard deviation of the magnitudes for common stars in f02 region with more than 3000 observations as a function of magnitude. The pink, blue, and green-dashed curves represent scintillation noise (calculated in Xiao et al. submitted), photon noise, and sky noise, respectively. The red-dotted lines represent a precision of 4\,mmag. The black-dashed curves and gray curves represent theoretically calculated magnitude errors, as a function of magnitude and median values, respectively. For MST COMS camera, dark current and readout noise are negligible (Zhang et al, submitted to this volume). The limiting magnitude for this case is about 18.5\,mag; more general scenarios will be presented in the data release paper (L. Mi, in preparation).}
    \label{fig:cali}
\end{figure}

\subsection{Astrometric Performance}
To quantitatively assess the precision of astrometry, we first cross-matched stars with aperture magnitude errors less than 0.01\,mag from more than 3000 images taken of the f02 region with Gaia DR3. Using the Gaussian fitting method, we derived the standard deviation of the cross-matching radius distribution between these sources and Gaia DR3. We tracked the nightly variation of this standard deviation and found it to be consistently around 80\,mas. Further, we analyzed the variation of this standard deviation over nearly 40 days, as shown in Fig.\,\ref{fig:astro3}, and observed that it remained consistently around 80\,mas. These findings indicate that the astrometric measurement precision is better than $0.1^{\prime\prime}$. 

Further improvements remain possible; for instance, within the $\texttt{SCAMP}$ process, we are considering replacing the existing polynomial fitting approach with a numerical smooth fitting method.

\subsection{Photometric Calibration Precision}
To validate the calibrated MST photometry, we performed two independent external verifications by combining it with LAMOST DR10 spectroscopic data, Gaia DR3, and recalibrated Pan-STARRS1 photometry \citep{2022AJ....163..185X,2023ApJS..268...53X}. First, we applied the spectroscopic-based stellar color regression method \citep{2015ApJ...799..133Y} and the color transformation approach with PS1 data to obtain the model magnitudes in the MST photometric system, respectively. Then, by comparing the differences between the SCR-based MST magnitudes and the MST magnitudes, and between the PS1-based MST magnitudes and the MST magnitudes, we derived the SCR- and PS1-based zero points, respectively. We found that the consistency between the SCR-/PS1-based zero points and the MST zero points remains stable over time, with the standard deviation of the zero point differences being less than 0.1\%. For more details, see Xiao et al. (to be submitted). The results indicate that the zero-point precision of the flux calibration is better than 0.1\%.

\subsection{Photometric Accuracy}
The final photometric uncertainty is dominated by both systematic and random errors and is related to the brightness of the source, as the random errors for faint sources are much larger than those for bright sources. Since the flux calibration is performed individually for each image, the MST's nearly 3000 repeated observations of almost a million sources in the f02 field provide a good choice for quantitatively analyzing the final photometric precision. We calculated the standard deviation of the magnitudes from nearly 3000 observations of about a million sources and plotted the variation of the standard deviation with brightness in Fig.\,\ref{fig:cali}. Additionally, we roughly estimated the theoretical SNR for sources of different brightness using the following equation, based on the observatory information, filter, camera, and telescope parameters:
\begin{eqnarray}
    \begin{aligned}
{\rm SNR_{total}}=\frac{N_{\rm star}}{[N_{\rm star}+N_{\rm sky}\cdot n_{\rm pix}+{\rm RN}^2\cdot n_{\rm pix}+N_{\rm dark}\cdot n_{\rm pix}]^\frac{1}{2}}~,
    \end{aligned}
\end{eqnarray}
\begin{eqnarray}
    \begin{aligned}
    {\rm SNR_{star}}=\sqrt{N_{\rm star}}~,~~~~~~{\rm SNR_{sky}}=N_{\rm star}/\sqrt{N_{\rm sky}\cdot n_{\rm pix}}~,
    \end{aligned}
\end{eqnarray}
\begin{eqnarray}
    \begin{aligned}
    n_{\rm pix}=\pi \times (0.5\times 2\cdot {\rm seeing / size})^2~,
    \end{aligned}
\end{eqnarray}
\begin{eqnarray}
    \begin{aligned}
    N_{\rm star/sky}=\eta \cdot \pi \times (\frac{D}{2})^2 \cdot \frac{F_{\rm star/sky}\cdot t}{h\cdot \nu} \cdot \frac{c}{\lambda^2}\cdot \Delta \lambda~,
    \end{aligned}
\end{eqnarray}
\begin{eqnarray}
    \begin{aligned}
    F_{\rm star/sky}=10^{-(m_{\rm star/sky}+48.6)/2.5}~.
    \end{aligned}
\end{eqnarray}
Where, $N_{\rm star}$ and $N_{\rm sky}$ are the total number of electrons from the star and the sky background per pixel, respectively. $N_{\rm pix}$ is the number of pixels within the aperture. $N_{\rm dark}$ and $RN$ are the number of electrons generated by the dark current and the readout noise per pixel, respectively, and neglected in this work. $\eta$ is the combined efficiency of the telescope’s optical system, quantum efficiency, filter, and atmospheric transmission. $\rm seeing$ is the observatory’s atmospheric seeing, $\rm size$ is the pixel scale, $D$ is the telescope diameter, $t$ is the exposure time, $c$ is the speed of light in vacuum, $h\nu$ is the energy of a photon, $\Delta \lambda$ and $\lambda$ are the bandwidth and central wavelength of the sensitivity, respectively. For example, for the MST G band, the parameters are as follows: $m_{\rm sky}=21$\,mag, $D=30$\,cm, $t=300$\,s, $\rm gain=0.25~e^{-}\,$ ADU$^{-1}$, $\rm seeing=2^{\prime\prime}$, $\rm size=0.86^{\prime\prime}\cdot$ pixel$^{-1}$, $\lambda=5400$\,\AA, $\Delta \lambda=900$\,\AA, and $\eta=0.32$.

From Fig.\,\ref{fig:cali}, it can be seen that at the bright end, the uncertainty is only 4\,mmag for $G~13$, slightly increasing to 10\,mmag for $G \sim 16$, and quickly grows to 40\,mmag toward the faint end at $G \sim 18$. This is in good agreement with the theoretical results. However, we observe a ``plateau'' in the magnitude scatter for repeated observations of the same source at both the bright and faint ends. This phenomenon goes beyond the scope of this paper and will be investigated further in future work.

\begin{figure}
    \centering
    \resizebox{\hsize}{!}{\includegraphics{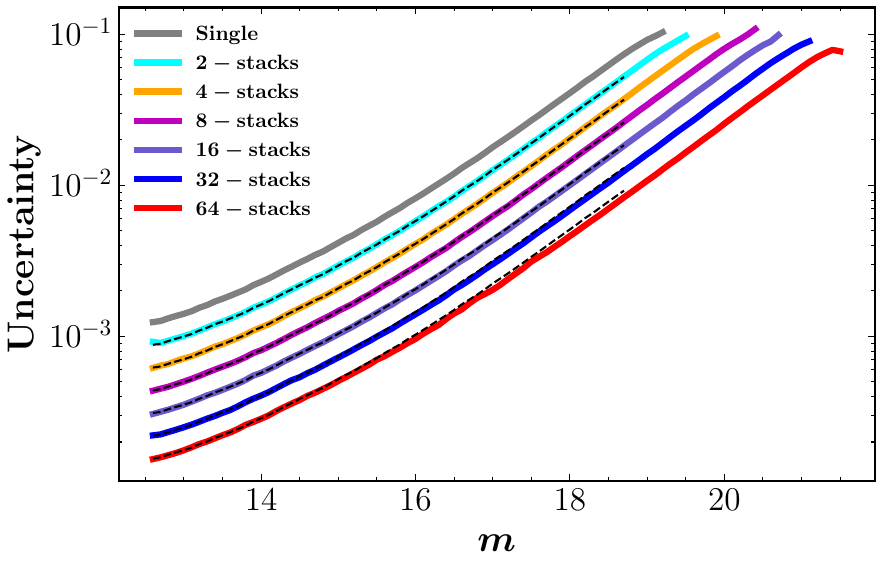}}
    \caption{An example showing the results of the image stack. Variation of magnitude error with MST $G$-band magnitude. The black dotted line represents the theoretical outcome obtained by multiplying the black solid line by $\sqrt{N}$, while the solid line denotes the median of the data points. $N$ is 2, 4, 8, 16, 32 and 64, which are labeled in the top left corner.}
    \label{fig:imsta}
\end{figure}

\subsection{Image and Catalog Stacks} \label{merging}
As a time-domain measurement project, MST requires high precision in light curve measurements. To reduce random errors and achieve high-precision light curves, one can stack catalogs or images while maintaining a consistent calibration accuracy. Catalog stacking involves phase-aligned stacking of light curves from single-image catalogs, while image stacking involves using $\texttt{SWarp}$ to combine images before performing precise measurements. The former is relatively easier to implement, while the latter can extend the limiting magnitude depth and observation completeness. A more detailed description of the image and catalog stacking processes will be found in Xiao et al. (in preparation).

\begin{figure}
    \centering
    \resizebox{\hsize}{!}{\includegraphics{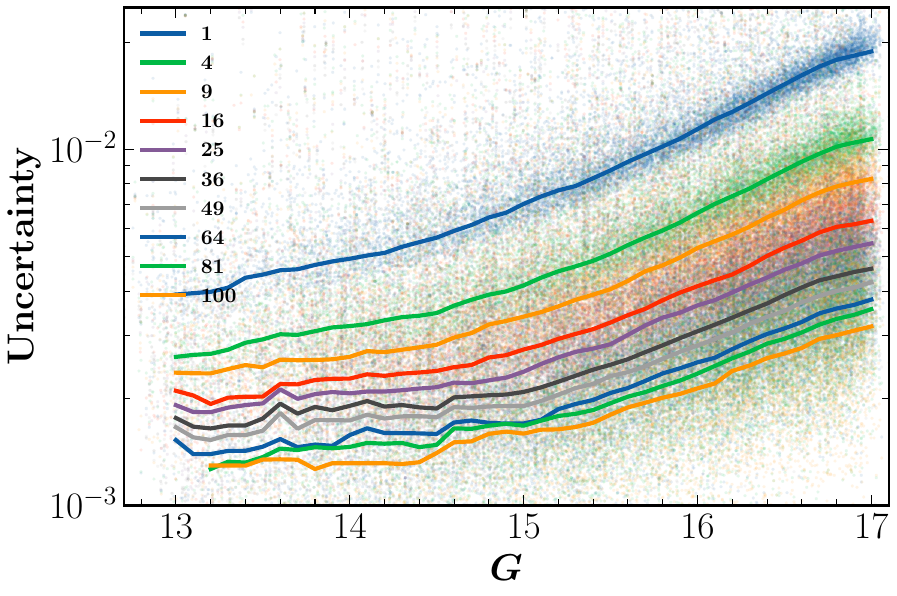}}
    \caption{Similar to Fig.\,\ref{fig:imsta}, but for the catalog stack. From top to bottom, the points represent the standard deviation of the magnitudes for common stars in f02 region after $1^2$ to $10^2$-stacks, respectively. The median values of the points are represented by the lines, respectively, as labeled in the top left corner.}
    \label{fig:casta}
\end{figure}

Here, we briefly test and evaluate the results of image and catalog stacking using MST $G$-band photometry, as shown in Fig.\,\ref{fig:imsta} and Fig.\,\ref{fig:casta}, respectively. When stacking 64 images, the $G$-band detection limit extends to 21\,mag, surpassing the single-image detection depth by 2\,mag. Additionally, the SNR increases significantly by a factor of $\sqrt{N}$ as the number of stacked images ($N$) increases up to 40. Furthermore, we utilize stars observed more than 1000 times to perform catalog stacking. We find that within 16-stacks of different observations of the same source, the consistency of magnitude measurements across repeated observations can be as low as 2\,mmag for bright sources; for bright stars, the consistency can reach up to 1.2\,mmag when utilizing 100-stacks. 

\subsection{Variable Source}
\begin{figure}
    \centering
    \resizebox{\hsize}{!}{\includegraphics{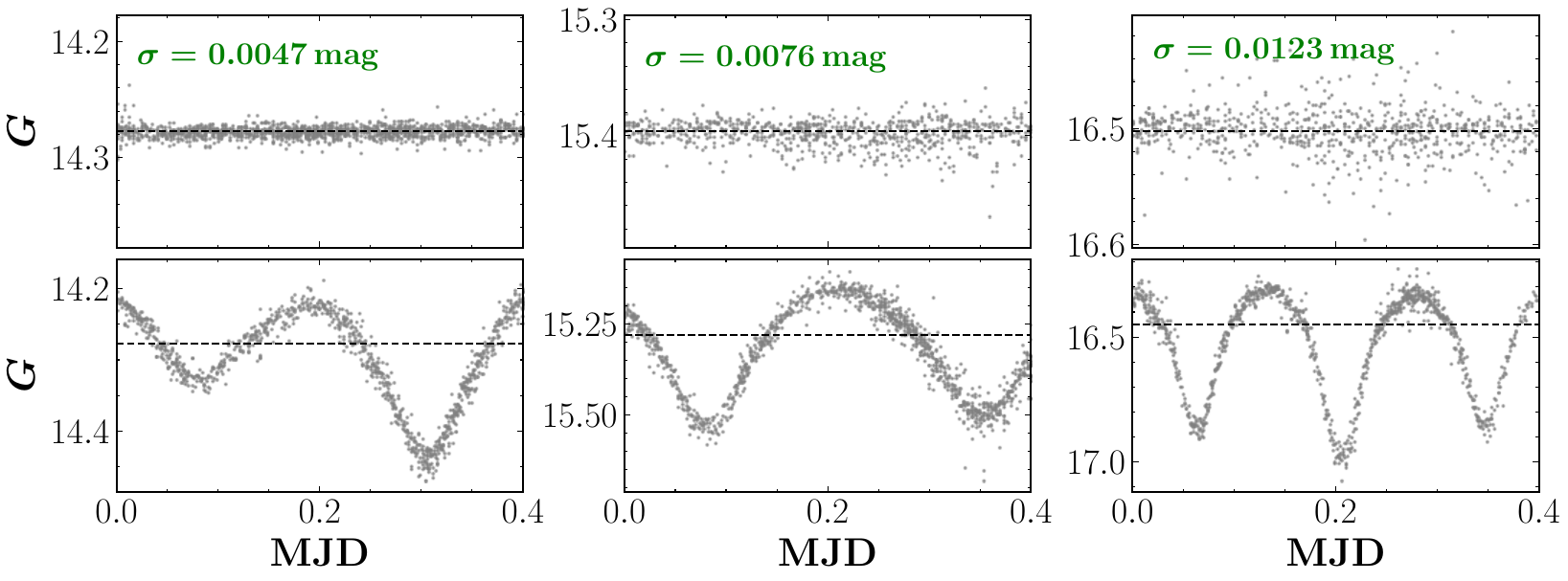}}
    \caption{{An example showing the light curves observed by MST2 ($G$-filter). Upper panels: The magnitude variations for 9 representative non-variable sources. Lower panels: Similar to the upper panels, but for 6 variable stars.}}
    \label{fig:variable}
\end{figure}

The study of variable sources plays an important role in astronomical research. 
Fig.\,\ref{fig:variable} shows the light curves of 3 non-variable sources and 3 variable sources with different magnitudes from MST2. We can see that as the sources become fainter, the scatter in their light curves increases, regardless of whether they are non-variable sources or variable sources. This is because the random errors increase as the sources become fainter. We calculated the standard deviation of the light curves for non-variable sources near 13\,mag, which is approximately 0.004\,mag, consistent with the results shown in Fig.\,\ref{fig:cali}.

\begin{figure}
    \centering
    \includegraphics[width=13cm]{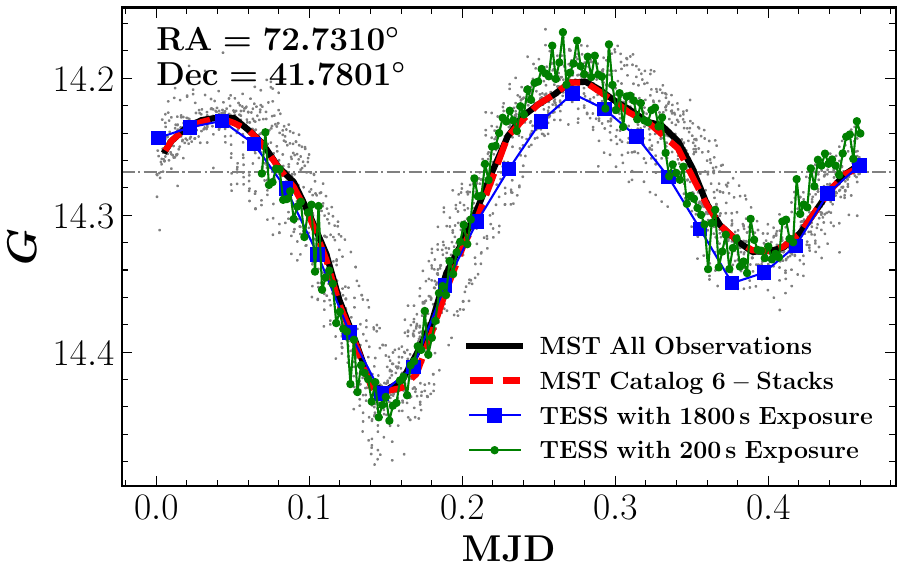}
    \caption{{The light curves of a variable source shown. The gray points represent MST $G$-band more than 3200 observations. The blue, green, black and red curves correspond to data from TESS with 1800\,s and 200\,s exposures, and the MST catalog after stacking with more than 3200 observations and 6-periods, respectively. The labels and the location of the source on the celestial sphere are indicated in the bottom-right and the upper-left corners, respectively.}}
    \label{fig:vstess}
\end{figure}

Additionally, Fig.\,\ref{fig:vstess} illustrates a comparison between the composite light curve derived from 6-stacks observations of a single source and the corresponding TESS light curve (\citealt{2020RNAAS...4..204H,2020RNAAS...4..206H,2021RNAAS...5..234K,2022RNAAS...6..236K}) at 1800\,s and 200\,s exposure.
Evaluating the pipeline's results demonstrates that the light curve obtained by MST is comparable to that obtained by TESS.

\subsection{Supernova}
\begin{figure} 
    \centering
    \includegraphics[width=14cm]{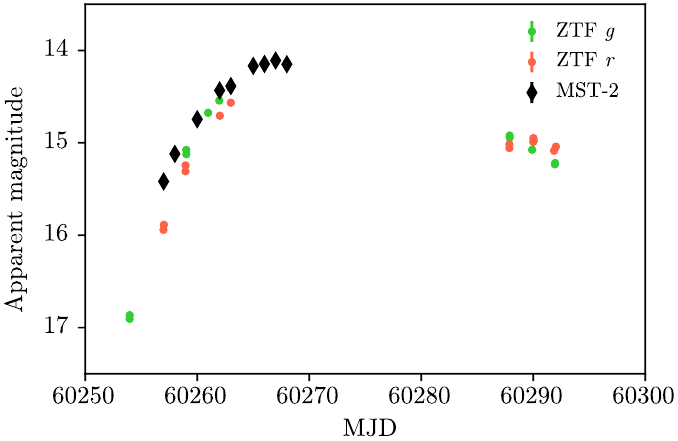}
    \caption{An example showing the results of supernova observed by MST2. ZTF photometry is also plotted for comparison. The legend is marked in the top right corner. 
  \label{fig:SN2023wrk}}
\end{figure}

Fig.\,\ref{fig:SN2023wrk} demonstrates an example of the light curve of a supernova observed as the experiment of searching for transient sources. Note that the photometric uncertainties are smaller than the size of markers. 
We also extract its light curve from ZTF Forced Photometry Service (ZFPs) \citep{2023Masci}, which shows good agreement with MST2 photometry based on the pipeline of this work.
As seen from Fig.\,\ref{fig:SN2023wrk}, it is difficult to cover all phases within a single survey; the MST or SiTian project is very important to sample light curves of transients from a different time zone.

\section{Summary} \label{secsum}

As a pathfinder of the SiTian project, MST at the Xinglong observation station houses three 30\,cm telescopes. Taking MST as an example, this work systematically introduces the data acquisition and management of a wide-field CMOS time-domain survey telescope, along with the development and application examples of a precise data processing pipeline tailored for wide-field instruments with CMOS detectors. This pipeline includes preprocessing, astrometry, aperture photometry, PSF photometry, and photometric calibration.

For the data acquisition and management, each MST telescope is connected to a standalone control computer (MST mini host), allowing the controller to wirelessly connect and issue ASCOM commands via the Observation Control Software (OCS). Telescope status is monitored in real time, and images are stored on the MST host and uploaded to the server for processing. Calibration frames (FLAT, DARK, BIAS) require manual intervention and are also uploaded after observation.

The evaluation of the performance of the MST precise data processing pipeline includes processing 3200 observational data sets collected by MST2 in the f02 region. This evaluation analyzes astrometric accuracy, photometric calibration precision, relative photometry and variability, variable sources, image stacking, and catalog merging, as detailed below:
\begin{enumerate}
    \item Comparison with Gaia DR3 reveals total two-axis astrometric residuals showcasing a stable total precision of less than 0.1$^{\prime\prime}$ over time.
    \item This work achieved an impressive flux calibration precison of approximately 1\,mmag in the MST zero-points. Calculation based on co-observed brighter stars indicates a standard deviation of agreement of about 4\,mmag.
    \item As a test, stacking 64 images enhances the $G$-band detection limit by more than 2\,mag compared to a single image, while the SNR increases significantly, scaling by a factor of $\sqrt{N}$ as the number of stacked images reaches 64. In particular, the catalog stacks indicates that the light curve of variable star after 6-stacks observations obtained by MST is comparable to that obtained by TESS.
    \item As an example, the light curve of supernova from ZTF Forced Photometry Service shows good agreement with MST2 photometry based the pipeline of this work.
\end{enumerate}

The results presented in this paper demonstrate that the performance of the CMOS used in MST is now comparable to that of CCD detectors, achieving similar photometric accuracy. This suggests that the MST CMOS could be considered as a viable alternative to CCDs in future large-scale time-domain surveys. However, CMOS also introduces errors that differ from those of CCDs, such as PS noise. This noise has a more significant impact on faint sources than on bright ones and should be given careful attention in future studies.

\clearpage
\section*{Acknowledgments}
We thank the anonymous referee for the helpful comments.
This work is supported by the National Key Basic R\&D Program of China via 2023YFA1608303 and the Strategic Priority Research Program of the Chinese Academy of Sciences (XDB0550103); the National Science Foundation of China 12422303, 12403024, 12222301, 12173007, and 12261141690; the Postdoctoral Fellowship Program of CPSF under Grant Number GZB20240731; the Young Data Scientist Project of the National Astronomical Data Center, and the China Post-doctoral Science Foundation No. 2023M743447. Z.X.N acknowledges support from the NSFC through grant No. 12303039 and No. 12261141690.

The SiTian project is a next-generation, large-scale time-domain survey designed to build an array of over 60 optical telescopes, primarily located at observatory sites in China. This array will enable single-exposure observations of the entire northern hemisphere night sky with a cadence of only 30-minute, capturing true color (gri) time-series data down to about 21 mag. This project is proposed and led by the National Astronomical Observatories, Chinese Academy of Sciences (NAOC). As the pathfinder for the SiTian project, the Mini-SiTian project utilizes an array of three 30 cm telescopes to simulate a single node of the full SiTian array. The Mini-SiTian has begun its survey since November 2022. The SiTian and Mini-SiTian have been supported from the Strategic Pioneer Program of the Astronomy Large-Scale Scientific Facility, Chinese Academy of Sciences and the Science and Education Integration Funding of University of Chinese Academy of Sciences.

\clearpage
\bibliographystyle{raa}
\bibliography{bibtex}

\begin{thebibliography}{25}
\providecommand\natexlab[1]{#1}
\providecommand\JournalTitle[1]{#1}

\bibitem[{Alarcon} {et~al.}(2023)]{2023PASP..135e5001A}
{Alarcon}, M.~R., {Licandro}, J., {Serra-Ricart}, M., {et~al.} 2023, \pasp,
  135, 055001

\bibitem[Bellm {et~al.}(2018)]{Bellm_2018}
Bellm, E.~C., Kulkarni, S.~R., Graham, M.~J., {et~al.} 2018, Publications of
  the Astronomical Society of the Pacific, 131, 018002

\bibitem[{Bertin}(2006)]{2006ASPC..351..112B}
{Bertin}, E. 2006, in Astronomical Society of the Pacific Conference Series,
  Vol. 351, Astronomical Data Analysis Software and Systems XV, ed.
  C.~{Gabriel}, C.~{Arviset}, D.~{Ponz}, \& S.~{Enrique}, 112

\bibitem[{Bertin}(2011)]{2011ASPC..442..435B}
{Bertin}, E. 2011, in Astronomical Society of the Pacific Conference Series,
  Vol. 442, Astronomical Data Analysis Software and Systems XX, ed. I.~N.
  {Evans}, A.~{Accomazzi}, D.~J. {Mink}, \& A.~H. {Rots}, 435

\bibitem[{Bertin} \& {Arnouts}(1996)]{1996A&AS..117..393B}
{Bertin}, E., \& {Arnouts}, S. 1996, \aaps, 117, 393

\bibitem[{Bertin} {et~al.}(2002)]{2002ASPC..281..228B}
{Bertin}, E., {Mellier}, Y., {Radovich}, M., {et~al.} 2002, in Astronomical
  Society of the Pacific Conference Series, Vol. 281, Astronomical Data
  Analysis Software and Systems XI, ed. D.~A. {Bohlender}, D.~{Durand}, \&
  T.~H. {Handley}, 228

\bibitem[{Borucki} {et~al.}(2010)]{2010Sci...327..977B}
{Borucki}, W.~J., {Koch}, D., {Basri}, G., {et~al.} 2010, Science, 327, 977

\bibitem[{Brandt}(2009)]{2009AAS...21332001B}
{Brandt}, J.~C. 2009, in American Astronomical Society Meeting Abstracts, Vol.
  213, American Astronomical Society Meeting Abstracts \#213, 320.01

\bibitem[Hough(1962)]{osti_4746348}
Hough, P.~V. 1962, 0

\bibitem[{Huang} {et~al.}(2024)]{2024ApJS..271...13H}
{Huang}, B., {Yuan}, H., {Xiang}, M., {et~al.} 2024, \apjs, 271, 13

\bibitem[{Huang} {et~al.}(2020{\natexlab{a}})]{2020RNAAS...4..204H}
{Huang}, C.~X., {Vanderburg}, A., {P{\'a}l}, A., {et~al.} 2020{\natexlab{a}},
  Research Notes of the American Astronomical Society, 4, 204

\bibitem[{Huang} {et~al.}(2020{\natexlab{b}})]{2020RNAAS...4..206H}
{Huang}, C.~X., {Vanderburg}, A., {P{\'a}l}, A., {et~al.} 2020{\natexlab{b}},
  Research Notes of the American Astronomical Society, 4, 206

\bibitem[{Kunimoto} {et~al.}(2022)]{2022RNAAS...6..236K}
{Kunimoto}, M., {Tey}, E., {Fong}, W., {et~al.} 2022, Research Notes of the
  American Astronomical Society, 6, 236

\bibitem[{Kunimoto} {et~al.}(2021)]{2021RNAAS...5..234K}
{Kunimoto}, M., {Huang}, C., {Tey}, E., {et~al.} 2021, Research Notes of the
  American Astronomical Society, 5, 234

\bibitem[{Lang} {et~al.}(2010)]{2010AJ....139.1782L}
{Lang}, D., {Hogg}, D.~W., {Mierle}, K., {Blanton}, M., \& {Roweis}, S. 2010,
  \aj, 139, 1782

\bibitem[{Liu} {et~al.}(2021)]{2021AnABC..93..628L}
{Liu}, J., {Soria}, R., {Wu}, X.-F., {Wu}, H., \& {Shang}, Z. 2021, Anais da
  Academia Brasileira de Ciencias, 93, 20200628

\bibitem[{Masci} {et~al.}(2023)]{2023Masci}
{Masci}, F.~J., {Laher}, R.~R., {Rusholme}, B., {et~al.} 2023, arXiv e-prints,
  arXiv:2305.16279

\bibitem[{Ofek} {et~al.}(2023)]{2023PASP..135f5001O}
{Ofek}, E.~O., {Ben-Ami}, S., {Polishook}, D., {et~al.} 2023, \pasp, 135,
  065001

\bibitem[{Ricker} {et~al.}(2015)]{2015JATIS...1a4003R}
{Ricker}, G.~R., {Winn}, J.~N., {Vanderspek}, R., {et~al.} 2015, Journal of
  Astronomical Telescopes, Instruments, and Systems, 1, 014003

\bibitem[{Stetson}(1990)]{1990PASP..102..932S}
{Stetson}, P.~B. 1990, \pasp, 102, 932

\bibitem[{Xiao} \& {Yuan}(2022)]{2022AJ....163..185X}
{Xiao}, K., \& {Yuan}, H. 2022, \aj, 163, 185

\bibitem[{Xiao} {et~al.}(2023)]{2023ApJS..268...53X}
{Xiao}, K., {Yuan}, H., {Huang}, B., {et~al.} 2023, \apjs, 268, 53

\bibitem[{Yuan} {et~al.}(2015)]{2015ApJ...799..133Y}
{Yuan}, H., {Liu}, X., {Xiang}, M., {et~al.} 2015, \apj, 799, 133

\bibitem[{Zhang} {et~al.}(2015)]{2015PASP..127.1292Z}
{Zhang}, J.-C., {Ge}, L., {Lu}, X.-M., {et~al.} 2015, \pasp, 127, 1292

\bibitem[{Zhang} {et~al.}(2020)]{2020PASP..132l5001Z}
{Zhang}, J.-C., {Wang}, X.-F., {Mo}, J., {et~al.} 2020, \pasp, 132, 125001

\end{thebibliography}

\end{document}